\documentclass{osa-article}

\journal{osajournal}

\articletype{Research Article}

\usepackage{lineno}

\usepackage{amsmath,algorithm,tabularx}
\usepackage[noend]{algpseudocode}
\usepackage[section]{placeins}
\usepackage{tabu} 
\usepackage{adjustbox}
\usepackage{ bbold }

\usepackage{graphicx}

\usepackage{subcaption} 
\usepackage{caption}
\usepackage{tikz}
\usepackage{amsthm}
\usepackage[english]{babel}
\usepackage{amsthm}
\usepackage{orcidlink}

\newtheorem{theorem}{Theorem}[section]
\newtheorem{lemma}[theorem]{Lemma}
\newtheorem{corollary}{Corollary}[theorem]
\newtheorem{definition}{Definition}[section]
\usepackage{xcolor}

\newcommand{\blue}[1]{\textcolor{cyan}{#1}}
\newcommand{\y}{\mathbf{y}}
\newcommand{\x}{\mathbf{x}}

\let\oldmathbf\mathbf
\let\oldboldsymbol\boldsymbol

\renewcommand{\mathbf}[1]{
  \ifcat\noexpand#1\relax 
    \oldboldsymbol{#1}
  \else
    \oldmathbf{#1}
  \fi
}

\renewcommand{\boldsymbol}[1]{
  \ifcat\noexpand#1\relax 
    \oldboldsymbol{#1}
  \else
    \oldmathbf{#1}
  \fi
}

\renewcommand{\eqref}[1]{\textup{{\normalfont~(\ref{#1}}\normalfont)}}

\newcommand{\A}{\mathbf{A}}

\begin{document}

\title{DDRM-PR: Fourier Phase Retrieval using Denoising Diffusion Restoration Models}

\author{Mehmet Onurcan Kaya\authormark{1,2}~\orcidlink{0009-0006-2606-3992} and Figen S. Oktem\authormark{3,*}~\orcidlink{0000-0002-7882-5120}}

\address{
\authormark{1}Dept. of Applied Mathematics and Computer Science, DTU, Lyngby, 2800, Denmark\\
\email{\authormark{2}monka@dtu.dk}
\authormark{3}Dept. of Electrical and Electronics Eng., METU, Ankara, 06800, Turkey\\
\email{\authormark{*}figeno@metu.edu.tr}
}

\begin{abstract*}
Diffusion models have demonstrated their utility as learned priors for solving various inverse problems. However, most existing approaches are limited to linear inverse problems. This paper exploits the efficient and unsupervised posterior sampling framework of Denoising Diffusion Restoration Models (DDRM) for the solution of nonlinear phase retrieval problem, which requires reconstructing an 
image from its noisy intensity-only measurements such as Fourier intensity. The approach combines the model-based alternating-projection methods with the DDRM to utilize pretrained unconditional diffusion priors for phase retrieval. The performance is demonstrated through both simulations and experimental data. Results demonstrate the potential of this approach for improving the alternating-projection methods as well as its limitations. 
\end{abstract*}

\section{Introduction}
\label{sec:introduction}

Fourier phase retrieval (PR) addresses the challenge of reconstructing a signal from 
its noisy Fourier magnitude measurements, a nonlinear inverse problem prevalent in diverse scientific and engineering domains such as crystallography, microscopy, astronomy, optical imaging, and speech processing \cite{Dong_2023, shechtman2015phase}. Despite its extensive applications, PR remains a formidable ill-posed inverse problem due to the inherent loss of phase information and the resultant non-convexity. Over the years, various algorithmic solutions have been proposed, each offering unique advantages and limitations.

Classical PR methods 
are based on alternating projection such as the
popular Gerchberg-Saxton (GS), error-reduction and hybrid input-output (HIO) algorithms. 
These methods alternate between space and measurement domains to impose the available constraints and intensity measurements through
projections. These classical methods are widely used because of their simple implementation, computational
efficiency, and applicability to different phase retrieval problems. Moreover, they  can empirically converge to a reasonably
good solution in various applications. However, they are
prone to suboptimal reconstructions due to noise amplification and being stuck in local minima. Advanced techniques leveraging semidefinite programming, sparse regularization, and global optimization have been developed to overcome these limitations, albeit at the expense of increased computational complexity or restrictive assumptions \cite{Fannjiang2020TheNO, stefanoqianpty, Maiden:17}.

Deep learning has 
enabled new capabilities for 
solving inverse problems in imaging, including phase retrieval \cite{Sinha:17}. Initially deep neural networks (DNNs) have achieved significant success in directly reconstructing images from measurements or enhancing initial estimates from classical methods~\cite{jin2017deep,Isil:19}. Model-based optimization schemes have also integrated deep priors within the plug-and-play framework~\cite{venkatakrishnan2013plug,chan2017,romano2017,pmlr-v80-metzler18a,Isil:20,isil2024,Wei2020TuningfreePP}. Nevertheless, existing deep learning solutions for PR are often hindered by the domain shift problem, lack of interpretability, and necessity for extensive training \cite{Wang_2024}.

Recently diffusion models have revolutionized the field of unconditional image generation, demonstrating superior performance across various tasks such as super-resolution, deblurring, inpainting, colorization, and compressive sensing~\cite{kawar2022denoising,dhariwal2021diffusion,Kawar2022JPEGAC}. 
These generative models stochastically denoise a sample to gradually produce the desired output,  conditioned on the measurements. 
The use of pretrained diffusion models allows for efficient and effective restoration without the need for specific training on individual degradation models, thereby offering great flexibility and adaptability in real-world applications. 
However, existing approaches are often applied to linear inverse problems. 

In this work, we exploit the efficient unsupervised posterior sampling framework of Denoising Diffusion Restoration Models (DDRM)~\cite{kawar2022denoising,Kawar2022JPEGAC} for the solution of nonlinear phase retrieval problem. Unlike existing methods, our approach does not require training; instead, it utilizes a pre-trained 
denoising diffusion model akin to plug-and-play methods. This enhances the practicality and ease of implementation, as it eliminates the need for additional training and complex parameter tuning.

By integrating the strengths of pretrained unconditional diffusion models with 
alternating projection techniques, our method provides an efficient solution to the challenging problem of phase retrieval, paving the way for further advancements in this field. The performance of the method is first demonstrated through simulations using both distortion and perceptual quality metrics 
to highlight its potential 
as well as its limitations for the Fourier phase retrieval problem. While this performance analysis is for real-valued images, our approach is applicable to complex-valued images as well as other type of phase retrieval data such as coded diffraction patterns and scattering measurements. This generalization is enabled by the flexibility of the framework allowing the integration of alternative denoisers without retraining as well as the adaptability of the alternating-projection methods to the specific measurement and signal constraints. 
To demonstrate this, the performance of the approach is also illustrated with experimental data acquired for imaging through random scattering media. The results highlight its capability to address phase retrieval tasks beyond the classical Fourier setting.

The paper is organized as follows. Section \ref{sec:background} provides an overview of the classical Fourier phase retrieval problem, while Section \ref{sec:relatedworks} reviews the related existing works. Our proposed approach is presented in Section \ref{sec:method}, followed by a comparative performance analysis 
in Section \ref{sec:results}. Lastly, Section \ref{sec:conclusion} summarizes our findings and outlines future research directions.

\section{Fourier Phase Retrieval Problem} 
\label{sec:background}

	In the classical Fourier phase retrieval problem, available measurements can be modeled as

	\begin{equation}\label{eq:poisson_pro}
	\mathbf{y^2} =   \mathbf{\vert {F}x \vert^2 + w}, \quad \quad \mathbf{w} \sim \mathcal{N}(\mathbf{0}, \alpha^2 \text{diag}(\vert \mathbf{{F}x} \vert^2)) 
	\end{equation}

where the measurement vector $\mathbf{y} \in \mathbb{R}^{m}$ denotes the lexicographical ordering of the $\sqrt{m} \times \sqrt{m}$ noisy Fourier intensity measurements. Likewise, the image vector $\mathbf{x} \in \mathbb{R}^{n}$ represents the lexicographical ordering of the $\sqrt{n} \times \sqrt{n}$ target image, which is assumed to be real-valued, non-negative, and of finite support. The matrix $\mathbf{{F}} \in \mathbb{C}^{m \times n}$ performs the oversampled $\sqrt{m} \times \sqrt{m}$-point discrete Fourier transform (DFT). The term $\mathbf{w} \in \mathbb{R}^{m}$ represents the measurement noise, and $\alpha$ is a scaling factor that adjusts the signal-to-noise ratio (SNR). The noise is generally Poisson-distributed but can often be approximated with a normal distribution~\cite{pmlr-v80-metzler18a}, as used here. Here the operator $\mathrm{diag}(\cdot)$ maps a vector to a diagonal matrix whose diagonal entries are given by the elements of the input vector.

For discrete real-valued signals in two or more dimensions with finite support, the Fourier intensity measurements at discrete frequencies 
can uniquely determine the unknown signal $\mathbf{x}$. To ensure uniqueness (aside from trivial ambiguities), for an image with support $\sqrt{n} \times \sqrt{n}$, it is required to measure the magnitude of its $\sqrt{m} \times \sqrt{m}$-point oversampled DFT with $\sqrt{m} \geq 2\sqrt{n} - 1$~\cite{hayes1982}. For simplicity, this work sets $m$ to $4n$.
Trivial ambiguities arise from the fact that there are 
operations that do not modify the Fourier magnitude, such as global phase shift, conjugate inversion, and spatial circular shift.

This mathematical problem is encountered across a wide range of real-world applications due to the equivalence of Fourier intensity to the autocorrelation function after inverse Fourier transform. 
The real-valued intensity distribution of the object is reconstructed from its Fourier intensity measurements or equivalently from its second-order correlations in various applications~\cite{Schulz1992image}. Examples include astronomical imaging to observe through turbulent atmosphere~\cite{1987_fienup_pr_astronomy}, laser-illuminated imaging to reconstruct diffuse objects without speckle noise~\cite{10.1117/12.7976761}, and imaging through random scattering media such as biological tissue~\cite{katz2014non}. There are also applications where the encountered phase retrieval problem can be approximated in this form. 

For instance, in 
crystallography, the formulation in Eq.~\eqref{eq:poisson_pro} is valid under the condition of low absorption \cite{miao2008extending}. When absorption is negligible, the real and imaginary components of the object's scattering and absorption coefficients can be simplified, allowing the object to be approximated as real-valued. These attributes highlight the 
importance of the considered Fourier phase retrieval problem in tackling challenging imaging scenarios across different domains.

\section{Related Works}
\label{sec:relatedworks}

\subsection{Classical Alternating Projection Methods for Phase Retrieval}
Alternating projection techniques have become fundamental tools for phase retrieval. One of the earliest and most well-known algorithms is the Gerchberg-Saxton (GS) algorithm~\cite{gs1978}, which reconstructs an unknown signal from magnitude constraints in the spatial and Fourier domains by performing iterative projections onto these non-convex constraint sets. A modification of the GS algorithm is the Error Reduction (ER) algorithm, which incorporates additional spatial domain constraints beyond just magnitude~\cite{fienup1978reconstruction}. A particularly significant and widely used method among alternating projection techniques is the Hybrid Input-Output (HIO) algorithm~\cite{fienup1982comparison}, which builds upon the principles of the ER algorithm.

In the HIO method, Fourier magnitude constraints and available space-domain constraints (such as support, non-negativity, and real-valuedness) are iteratively applied, similar to the ER algorithm. However, the key distinction is that HIO does not force the iterates to strictly satisfy the constraints at every step. Instead, it uses the iterates to progressively guide the algorithm towards a solution that meets the constraints~\cite{fienup1982comparison}. The HIO iterations are mathematically expressed as follows:
\begin{equation}\label{eq:prnetttthio2}
\begin{aligned}
\mathbf{x}_{k+1}[n] = \left\{ \begin{array}{rcl}
\mathbf{u}_k[n] & \text{for} & n \notin \gamma \\
\mathbf{x}_{k}[n] - \beta \mathbf{u}_k[n] & \text{for} & n \in \gamma \\
\end{array}\right.
\end{aligned}
\end{equation}
where
\begin{equation}\label{eq:prnetttthio1}
\begin{aligned}
\mathbf{u}_k = \mathbf{F}^{-1}\left\{\mathbf{y} \odot \frac{\mathbf{F} \mathbf{x}_k}{\vert \mathbf{F} \mathbf{x}_k \vert}\right\}.
\end{aligned}
\end{equation}
Here $\mathbf{x}_k \in \mathbb{R}^{m}$ represents the reconstruction at the $k^{\text{th}}$ iteration, $\mathbf{F^{-1}}$ denotes the inverse DFT matrix, $\odot$ is element-wise multiplication, $\beta$ is a constant parameter (commonly set to 0.9), and $\gamma$ is the set of indices $n$ where $\mathbf{u}_k[n]$ fails to meet the spatial domain constraints~\cite{fienup1982comparison}.

Despite the lack of a comprehensive theoretical understanding of its convergence behavior, HIO method has been empirically observed to converge to reasonably good solutions in a wide range of applications. However, the reconstructions produced by HIO can contain artifacts and errors due to getting trapped in local minima or the amplification of noise in the solution~\cite{shechtman2015phase, marchesini2007invited}. To address these limitations, numerous variations and enhancements of the HIO method have been proposed, aiming to improve its reconstruction performance and reliability~\cite{stefanoqianpty, Maiden:17}.

\subsection{Diffusion Models}
Diffusion models approximate a data distribution $q(x)$ by learning a model distribution $p_\theta(x)$ from samples. These generative models possess a Markov chain structure, represented as $\mathbf{x}_T \rightarrow \mathbf{x}_{T-1} \rightarrow \ldots \rightarrow \mathbf{x}_1 \rightarrow \mathbf{x}_0$, where $\mathbf{x}_t \in \mathbb{R}^n$, with their joint distribution given by~\cite{kawar2022denoising}
\begin{equation}
p_\theta\left(\mathbf{x}_{0: T}\right)=p_\theta^{(T)}\left(\mathbf{x}_T\right) \prod_{t=0}^{T-1} p_\theta^{(t)}\left(\mathbf{x}_t \mid \mathbf{x}_{t+1}\right).
\end{equation}
Once $\mathbf{x}_{0: T}$ is generated, only $\mathbf{x}_0$ is retained as the sample from the generative model. A fixed 
variational inference distribution is utilized to train the diffusion model:
\begin{equation}
q\left(\mathbf{x}_{1: T} \mid \mathbf{x}_0\right)=q^{(T)}\left(\mathbf{x}_T \mid \mathbf{x}_0\right) \prod_{t=0}^{T-1} q^{(t)}\left(\mathbf{x}_t \mid \mathbf{x}_{t+1}, \mathbf{x}_0\right)
\end{equation}
This approach results in an evidence lower bound (ELBO) for the maximum likelihood objective. For diffusion models where both $p_\theta^{(t)}$ and $q^{(t)}$ are selected as conditional Gaussian distributions for all $t<T$, $\mathbf{x}_t$ can be viewed as corrupted version of $\mathbf{x}_0$ with Gaussian noise since $q\left(\mathbf{x}_t \mid \mathbf{x}_0\right)$ becomes a Gaussian distribution with known mean and covariance. Mathematically, this can be expressed as $q\left(\mathbf{x}_t \mid \mathbf{x}_0\right)=\mathcal{N}\left(\sqrt{\alpha_t} \mathbf{x}_0,\left(1-\alpha_t\right) \boldsymbol{I}\right), \; \forall t \in[1, T]$. 
As a result, the ELBO objective simplifies into the following denoising autoencoder objective, as detailed in \cite{li2023diffusion}:
\begin{equation}
\sum_{t=1}^T \gamma_t \mathbb{E}_{\left(\mathbf{x}_0, \mathbf{x}_t\right) \sim q\left(\mathbf{x}_0\right) q\left(\mathbf{x}_t \mid \mathbf{x}_0\right)}\left[\left\|\mathbf{x}_0-f_\theta^{(t)}\left(\mathbf{x}_t\right)\right\|_2^2\right]
\end{equation}
Here $\gamma_{1: T}$ are positive coefficients 
and $f_\theta^{(t)}$ represents a neural network with parameters $\theta$, whose purpose is to recover noiseless $\mathbf{x}_0$ from its noisy observation $\mathbf{x}_t$.

While the theoretical formulation of diffusion models emphasizes the recovery of $\mathbf{x}_0$ from noisy observations $\mathbf{x}_t$, practical implementations often focus on parametrizing $f_\theta^{(t)}$ to predict the noise added during the forward process. This approach, commonly referred to as noise estimation, has been found to simplify the learning task for neural networks and improve empirical performance, as shown in \cite{ho2020denoising, saharia2022photorealistic}. Moreover, advancements in diffusion models, such as working in latent spaces or integrating noise prediction with hybrid objectives, have further enhanced their efficiency and versatility \cite{rombach2022high}. These practical considerations underline the flexibility of diffusion models in adapting their parametrizations to different objectives while preserving their theoretical underpinnings.

\subsubsection{Denoising Diffusion Restoration Models (DDRM)}
DDRM have been proposed as a versatile solution for addressing linear inverse problems in both noisy and noiseless contexts. Specifically, DDRM functions as a general solver for the inverse problems with the following forward model: $\mathbf{y} = \boldsymbol{H}\mathbf{x} + \mathbf{z}$ where $\mathbf{z} \sim \mathcal{N}(\mathbf{0}, \sigma_{\mathbf{y}}^2 \mathbf{I})$. It is defined by a Markov chain model conditioned on $\mathbf{y}$~\cite{kawar2022denoising}:
\begin{equation}
p_\theta\left(\mathbf{x}_{0:T} \mid \mathbf{y}\right) = p_\theta^{(T)}\left(\mathbf{x}_T \mid \mathbf{y}\right) \prod_{t=0}^{T-1} p_\theta^{(t)}\left(\mathbf{x}_t \mid \mathbf{x}_{t+1}, \mathbf{y}\right)
\end{equation}
where \(\mathbf{x}_0\) represents the final output. To train the diffusion model, a variational distribution conditioned on $\mathbf{y}$ is considered. 

DDRM employs a procedure that leverages a pretrained unconditional diffusion model 
similar to pretrained denoisers used in plug-and-play methods. Notably, this approach eliminates the necessity for additional training. The authors demonstrate that, under specific conditions, the solution obtained by training a conditional diffusion model is equivalent to that derived from using a pretrained unconditional diffusion model in conjunction with the DDRM procedure. Consequently, this equivalence enables to effectively solve any linear inverse problem by utilizing a pretrained diffusion model, and hence simplifies the implementation while enhancing the practicality of the method. 

The core concept of DDRM is to utilize the singular value decomposition (SVD) of the matrix \(\boldsymbol{H}\), and transform both the target variable \(\x\) and the corrupted observation \(\y\) into a common spectral space. For each index of this spectral space, DDRM performs differently based on the provided information from \(\y\), as indicated by the singular values. For indices corresponding to non-zero singular values, DDRM performs denoising on $\y$, while for those associated with zero singular values, it undertakes regular unconditional generation without using $\y$. This approach explicitly accounts for measurement noise, thereby enhancing the robustness and accuracy of the restoration process~\cite{kawar2022denoising}.

In particular, the original form of DDRM

uses the singular value decomposition, i.e., $\boldsymbol{H}=\mathbf{U \Sigma V^T}$ with $\mathbf{U U^T}=\mathbf{I}$ and $\mathbf{V V^T}=\mathbf{I}$, to perform the diffusion in the spectral space:
\begin{eqnarray}
\mathbf{\bar{x}}_t &=& \mathbf{V^T x_t} \label{transformation1}\\
\mathbf{\bar{y}} &=& \mathbf{\Sigma^\dagger U^T y} \label{transformation2}
\end{eqnarray}
By defining $\mathbf{x}_{\theta,t} = \mathbf{f}_{\theta}^{(t+1)}(\mathbf{x}_{t+1})$ as a denoiser with parameters $\theta$ and letting $\mathbf{\bar{x}}_{\theta, t} = \mathbf{V^T}\mathbf{x}_{\theta,t}$, the DDRM sampling procedure is given as follows~\cite{kawar2022denoising}:
\begin{equation}
\begin{aligned}
p_\theta^{(T)}\left(\overline{\mathbf{x}}_T^{(i)} \mid \mathbf{y}\right)= & \begin{cases}\mathcal{N}\left(\overline{\mathbf{y}}^{(i)}, \sigma_T^2-\frac{\sigma_{\mathbf{y}}^2}{s_i^2}\right) & \text { if } s_i>0 \\
\mathcal{N}\left(0, \sigma_T^2\right) & \text { if } s_i=0\end{cases} \\
p_\theta^{(t)}\left(\overline{\mathbf{x}}_t^{(i)} \mid \mathbf{x}_{t+1}, \mathbf{y}\right)= & \begin{cases}\mathcal{N}\left(\overline{\mathbf{x}}_{\theta, t}^{(i)}+\sqrt{1-\eta^2} \sigma_t \frac{\overline{\mathbf{x}}_{t+1}^{(i)}-\overline{\mathbf{x}}_{\theta, t}^{(i)}}{\sigma_{t+1}}, \eta^2 \sigma_t^2\right) \\ \hfill \text { if } s_i=0 \\[10pt]
\mathcal{N}\left(\overline{\mathbf{x}}_{\theta, t}^{(i)}+\sqrt{1-\eta^2} \sigma_t \frac{\overline{\mathbf{y}}^{(i)}-\overline{\mathbf{x}}_{\theta, t}^{(i)}}{\sigma_{\mathbf{y}} / s_i}, \eta^2 \sigma_t^2\right) \\ \hfill \text { if } \sigma_t<\frac{\sigma_{\mathbf{y}}}{s_i} \\[10pt]
\mathcal{N}\left(\left(1-\eta_b\right) \overline{\mathbf{x}}_{\theta, t}^{(i)}+\eta_b \overline{\mathbf{y}}^{(i)}, \sigma_t^2-\frac{\sigma_{\mathbf{y}}^2}{s_i^2} \eta_b^2\right) \\ \hfill \text { if } \sigma_t \geq \frac{\sigma_{\mathbf{y}}}{s_i}\end{cases}
\end{aligned}
\end{equation}
where the $i$th singular value of $\boldsymbol{H}$ is $s_i$, the $i$-th index of any vector $\mathbf{x}$ is denoted by $\mathbf{x}^{(i)}$, and $\eta \in (0, 1]$ is a hyperparameter that changes the variance of the transitions. 
Moreover, $\mathbf{x}_{\theta,t} = \mathbf{f}_{\theta}^{(t+1)}(\mathbf{x}_{t+1})$ is trained with the regular unconditional diffusion process due to the conjugate variational distribution satisfying similar properties:
\begin{equation}
q\left(\mathbf{x}_t \mid \mathbf{x}_0\right)=\mathcal{N}\left(\mathbf{x}_0, \sigma_t^2 \boldsymbol{I}\right) \quad \text{with} \quad 0=\sigma_0 < \sigma_1 < .... < \sigma_T.
\end{equation}

\section{DDRM for Phase Retrieval}
\label{sec:method}

In order to use the DDRM framework for the nonlinear inverse problem of JPEG artifact correction, its simplified form for the linear and noiseless case has been provided~\cite{Kawar2022JPEGAC}. Our approach is also based on this simplified form of DDRM. 

\begin{theorem}[Simplified form of DDRM]
    Under a noiseless setting, i.e., $\sigma_\mathbf{y} = 0$, the overall DDRM process for linear inverse problems can be expressed as 
\begin{equation}
\begin{aligned}
\mathbf{x}_t^{\prime}=&\mathbf{x}_{\theta,t}-\boldsymbol{H}^{\dagger} \boldsymbol{H} \mathbf{x}_{\theta,t}+\boldsymbol{H}^{\dagger} \mathbf{y} \\
\mathbf{x}_t=&\sqrt{\alpha_t}\left(\eta_b \mathbf{x}_t^{\prime}+\left(1-\eta_b\right) \mathbf{x}_{\theta,t}\right) \\
&+\sqrt{1-\alpha_t}\left(\eta \mathbf{\epsilon}_t+(1-\eta) \mathbf{\epsilon}_\theta^{(t+1)}\left(\mathbf{x}_{t+1}\right)\right)
\end{aligned}
\end{equation}
where $\boldsymbol{H}^{\dagger}$ represents the Moore-Penrose pseudo-inverse of $\boldsymbol{H}$, i.e.  $\boldsymbol{H}^\dagger=\mathbf{V \Sigma^\dagger U^T}$. The term \( \mathbf{x}_{\theta,t} = f_\theta^{(t+1)}\left(\mathbf{x}_{t+1}\right) \) corresponds to the output of the denoising model at step \( t+1 \), while \( \mathbf{\epsilon}_\theta^{(t+1)}\left(\mathbf{x}_{t+1}\right) = \frac{\mathbf{x}_{t+1} - \sqrt{\alpha_{t+1}} \mathbf{x}_{\theta,t}}{\sqrt{1-\alpha_{t+1}}} \) denotes the estimated noise value. The constants \( \eta \) and \( \eta_b \) are hyperparameters defined by the user from the interval $(0,1]$, and \( \mathbf{\epsilon}_t \sim \mathcal{N}(\mathbf{0}, \boldsymbol{I}) \) is a vector drawn from a standard Gaussian distribution.

\end{theorem}

This simplified form has been provided in \cite{Kawar2022JPEGAC} without a proof. We provide a  proof in Appendix \ref{appendix:proofs}.

\subsection{DDRM for Fourier Phase Retrieval}
Although DDRM is derived for a linear operator $\boldsymbol{H}$, the provided simplified form suggests that it can also be used for nonlinear inverse problems where an operator exists to approximate the pseudo-inverse
~\cite{Kawar2022JPEGAC}. For the Fourier phase retrieval problem, we use the HIO algorithm for this purpose since in the noise-free case
\begin{itemize}
    \item applying HIO after taking the Fourier magnitude provides the original image subject to trivial ambiguities (similar to $\boldsymbol{H}^\dagger \boldsymbol{H} \mathbf{x}$ being close to $\mathbf{x}$ in the least-squares sense),
    \item computing the Fourier magnitude after applying the HIO method to the noiseless measurements provides the same measurement (similar to $\boldsymbol{H} \boldsymbol{H}^\dagger \boldsymbol{H} = \boldsymbol{H}$ i.e. applying the pseudo-inverse does not alter the noiseless measurement).
\end{itemize}

Based on these observations, we propose DDRM-PR for the Fourier phase retrieval problem as follows:
\begin{equation} \label{ourApproach}
\begin{aligned}
\mathbf{x}_t^{\prime}=&f_\theta^{(t+1)}\left(\mathbf{x}_{t+1}\right)- \text{HIO}( \mathbf{\vert F} f_\theta^{(t+1)}
\left(\mathbf{x}_{t+1}
\right) \vert )
+
\text{RandomInit}( \mathbf{y} ) \\
\mathbf{x}_t=&\sqrt{\alpha_t}\left(\eta_b \mathbf{x}_t^{\prime}+\left(1-\eta_b\right) f_\theta^{(t+1)}\left(\mathbf{x}_{t+1}\right)\right)\\
&+\sqrt{1-\alpha_t}\left(\eta \mathbf{\epsilon}_t+(1-\eta) \mathbf{\epsilon}_\theta^{(t+1)}\left(\mathbf{x}_{t+1}\right)\right)
\end{aligned}
\end{equation}
Here, RandomInit represents to the HIO initialization procedure proposed in the prDeep paper~\cite{pmlr-v80-metzler18a}. It involves executing the HIO method for $s=50$ iterations using $m=50$ different random initializations. Then the reconstruction with the smallest residual is selected for a final HIO run of $n=1000$ iterations. For $\text{HIO}( \mathbf{\vert F} f_\theta^{(t+1)}
\left(\mathbf{x}_{t+1}
\right) \vert )$, the algorithm is run for $k=100$ iterations from the initialization $\mathbf{x}_{t+1}$.
Furthermore, to ensure consistent performance, we generate $N=8$ independent outputs for each input and use the averaged image obtained from these outputs.
To set the hyperparameters including $\eta$, $\eta_b$, uniformly-spaced diffusion steps $t$, initial timestep $T_{init}$, and the number of averaged samples $N$, linear grid search is performed for the optimal choice.

Our method exploits pretrained unconditional diffusion models which offer several practical advantages. First of all, the pretrained models are developed using large and diverse image datasets, capturing a wide range of features that are crucial for effective denoising and reconstruction. The pretrained model acts as a strong prior, facilitating accurate reconstruction by refining noisy inputs iteratively. This integration also bypasses the need for retraining specific to the phase retrieval task, making the method more accessible and easier to implement in various settings.

\subsection{Extension to Other Phase Retrieval Problems} \label{section:extension}

For a general phase retrieval problem, the Fourier transform matrix $\mathbf{F}$ 
in Eq.~\eqref{eq:poisson_pro} should be replaced with the 
corresponding system matrix $\mathbf{A}$, which may or may not be invertible. The straightforward extension of the DDRM-PR approach to this general case can be given as follows:
\begin{equation} \label{extendedApproach}
\begin{aligned}
\mathbf{x}_t^{\prime}=&f_\theta^{(t+1)}\left(\mathbf{x}_{t+1}\right)- \text{AP}( \mathbf{\vert A} f_\theta^{(t+1)}
\left(\mathbf{x}_{t+1}
\right) \vert )
+
\text{RandomInit}( \mathbf{y} ) \\
\mathbf{x}_t=&\sqrt{\alpha_t}\left(\eta_b \mathbf{x}_t^{\prime}+\left(1-\eta_b\right) f_\theta^{(t+1)}\left(\mathbf{x}_{t+1}\right)\right)\\
&+\sqrt{1-\alpha_t}\left(\eta \mathbf{\epsilon}_t+(1-\eta) \mathbf{\epsilon}_\theta^{(t+1)}\left(\mathbf{x}_{t+1}\right)\right)
\end{aligned}
\end{equation}
Here, AP represents an appropriate alternating-projection (i.e. GS-type) method for the considered phase retrieval task, whereas RandomInit represents the same initialization procedure as before with the HIO method replaced with the chosen AP method.  

In the case that there are no space-domain constraints (such as support, real-valuedness, nonnegativity, etc.), AP method enforces only the measurement constraints through the projection $\mathbf{A}^{-1}\left\{\mathbf{y} \odot \frac{\mathbf{A} \mathbf{x}_k}{\vert \mathbf{A} \mathbf{x}_k \vert}\right\}$ for an invertible $\mathbf{A}$. If $\A$ is non-invertible, similar to the well-known extensions of alternating-projection methods to the general phase retrieval problems~\cite{netrapalli2013phase, metzler2017coherent, chandra2019phasepack}, 
the pseudoinverse $\A^\dag$ is used (instead of matrix inverse) to go back from the measurement space to the object space. That is, measurement constraints are enforced through $\mathbf{A}^\dag\left\{\mathbf{y} \odot \frac{\mathbf{A} \mathbf{x}_k}{\vert \mathbf{A} \mathbf{x}_k \vert}\right\}$ for a non-invertible $\A$. 
For better computational efficiency, explicit calculation of the pseudoinverse can be avoided. Instead, equivalent least squares problem can be solved using computationally cheap iterative methods such as the conjugate gradient method~\cite{shewchuk1994introduction}.

Note also that pre-trained diffusion models used in our approach should be appropriately chosen for the target image type. In our implementations, we assumed a natural image distribution for the real-valued target image and employed a pre-trained diffusion model trained on natural images.  When dealing with phase-only objects~\cite{Sinha:17} which are real-valued but constrained to have values within a range different than natural images (such as $[0, 2\pi]$), a pre-trained diffusion model trained on such phase objects is needed.

Moreover, to recover complex-valued images (rather than real-valued ones), generally two different strategies are used. The first involves applying the pipeline separately to the amplitude and phase components, or to the real and imaginary parts of the complex-valued signal. This approach leverages the real-valued nature of these components to continue to use conventional real-valued DNNs. The second approach is to train a diffusion model specifically designed for complex-valued inputs and outputs using complex-valued neural networks \cite{lee2022complex, bassey2021survey}. While standard diffusion models typically employ real-valued U-Nets, extending them to handle complex-valued data is feasible \cite{choi2018phase}. Having access to images sampled from the target distribution of the complex-valued signal, we can utilize a pretrained diffusion model trained on such samples. Obviously this also requires to have a large dataset for training. If the real or imaginary components, or the amplitude or phase parts of the complex-valued image match the natural image distribution, the same pre-trained diffusion model used in this paper can simply be used. By using either of these two strategies, our method can accommodate the reconstruction of complex-valued signals in a wide range of phase retrieval applications. 

Hence our method can be applied to diverse phase retrieval problems beyond the classical Fourier PR setting with proper modifications.

\section{Results}
\label{sec:results}

\subsection{Performance Analysis with Simulated Data}

To evaluate the effectiveness of the proposed method, we first used the publicly available CelebA-HQ dataset \cite{karras2018}
that contains pictures of imaginary celebrities produced by a generative adversarial network (GAN) at a resolution of 256x256 pixels. The choice of RGB images is twofold: firstly, RGB images tend to reveal the reconstruction artifacts 
more clearly, and secondly, the pretrained diffusion models we employed are 
for RGB images. Each color channel (Red, Green, and Blue) is processed separately by the HIO algorithm to ensure that color information is preserved and accurately reconstructed.
After applying the HIO algorithm to each channel, we calculated the Peak Signal-to-Noise Ratio (PSNR) values to correct for conjugate inversion ambiguity. This step is crucial for ensuring the accuracy of the phase retrieval process, as it aligns the reconstructed images more closely with the ground truth.

The image quality metrics used to assess the quality of the reconstructed images are PSNR, Structural Similarity Index (SSIM) \cite{wang2004image}, and Learned Perceptual Image Patch Similarity (LPIPS) \cite{zhang2018unreasonable}. 
PSNR quantitatively measures the reconstruction fidelity, whereas SSIM and LPIPS emphasize the perceptual quality of the reconstructed images. While phase retrieval is primarily a quantitative problem, diffusion models are renowned for their ability to produce visually sharp and realistic outputs, often outperforming MMSE- or MAP-based methods that typically yield overly smooth or blurry reconstructions. SSIM, though a distortion metric, correlates well with human visual perception and offers insights into structural similarity. LPIPS, on the other hand, directly measures perceptual similarity by comparing deep feature representations, which makes it particularly useful for assessing the visual realism of the generated images. While our primary focus remains on achieving low distortion metrics like PSNR, the inclusion of perceptual metrics highlights the dual capability of our approach to maintain both quantitative accuracy and perceptual quality. This dual evaluation underscores the power of diffusion models in generating reconstructions that have high fidelity, less artifacts and also visual coherence, demonstrating a clear advantage over conventional phase retrieval techniques.

To evaluate the performance of the proposed method, we conducted experiments under various noise levels. We applied the proposed method to a diverse set of test images from the CelebA-HQ dataset with varying levels of measurement noise, 
i.e., with different values of $\alpha$. For each noise level, we generated multiple reconstructions and computed the average PSNR, SSIM, and LPIPS as given in Table \ref{table:table1}. 

\begin{table}[htb!]
	\centering
	\caption{\bf The average reconstruction performances for CelebA-HQ test images.}
	\small    
        \begin{adjustbox}{max width=\columnwidth}
     
		\begin{tabu}{@{\extracolsep{4pt}}ccccccccccccc@{}}
			\hline
&\multicolumn{3}{c}{$\alpha = 0.5$} &\multicolumn{3}{c}{$\alpha = 1$}&\multicolumn{3}{c}{$\alpha = 2$}&\multicolumn{3}{c}{$\alpha = 3$} \\
\cline{2-4} \cline{5-7} \cline{8-10} \cline{11-13}
Methods &PSNR $\uparrow$ &SSIM $\uparrow$&LPIPS $\downarrow$&PSNR $\uparrow$&SSIM $\uparrow$&LPIPS $\downarrow$&PSNR $\uparrow$&SSIM $\uparrow$&LPIPS $\downarrow$&PSNR $\uparrow$&SSIM $\uparrow$&LPIPS $\downarrow$
\\
\hline
HIO Stage&28.74&0.82&0.14&27.57&0.74&0.21&25.27&0.65&0.34&24.00&0.58&0.43\\
DDRM-PR&29.13&0.87&0.13&28.45&0.84&0.15&26.59&0.79&0.23&25.73&0.76&0.27\\
   
			\hline					
	\end{tabu}
     \end{adjustbox}

\label{table:table1}
 
\end{table}

Our method demonstrated superior performance compared to 
HIO for all evaluated metrics. The PSNR values indicated that our method is capable of suppressing various artifacts and noise while preserving important image details. The SSIM scores showed that the structural integrity of the images was well-maintained, and the high LPIPS values confirmed the perceptual quality of the reconstructions. We observed that the use of multiple independent outputs and averaging also help to enhance the stability and reliability of the results.

\begin{figure}[htb!]
    \centering

    \includegraphics[width=\columnwidth]{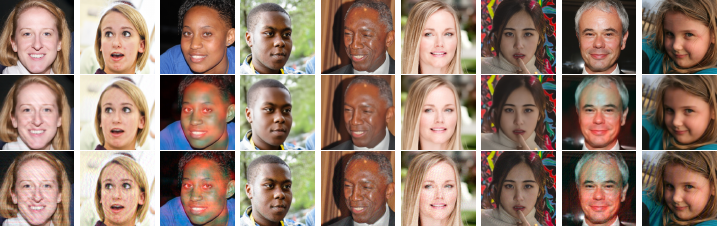}
    
\caption{Ground-truth test images (top row), reconstructions using the developed approach (middle row), and HIO initialization results (bottom row) for the case with parameters: $\alpha=0.5$, $N=1$, $\eta=0.15$, $\eta_b=0.20$, $t=15$, and $T_{init}=350$.}
\label{fig:fig1}
\end{figure}

\begin{figure}[htb!]
    \centering

    \includegraphics[width=\columnwidth]{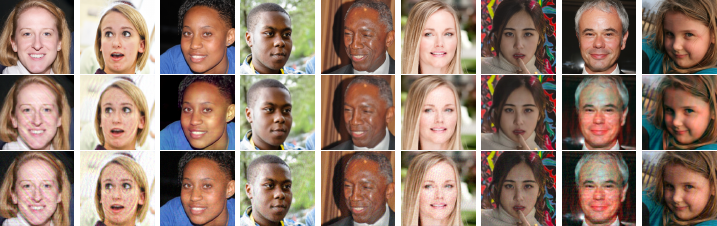}

\caption{Ground-truth test images (top row), reconstructions using the developed approach (middle row), and HIO initialization results (bottom row) for the case with parameters: $\alpha=1$, $N=1$, $\eta=0.25$, $\eta_b=0.22$, $t=30$, and $T_{init}=400$.}
\label{fig:fig2}
\end{figure}

\begin{figure}[htb!]
    \centering

    \includegraphics[width=\columnwidth]{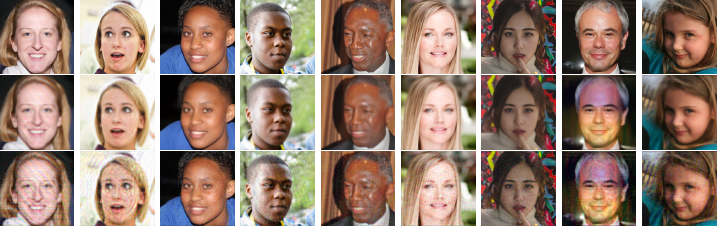}

\caption{Ground-truth test images (top row), reconstructions using the developed approach (middle row), and HIO initialization results (bottom row) for the case with parameters: $\alpha=2$, $N=1$, $\eta=0.25$, $\eta_b=0.18$, $t=15$, and $T_{init}=400$.}
\label{fig:fig3}
\end{figure}

\begin{figure}[htb!]
    \centering

    \includegraphics[width=\columnwidth]{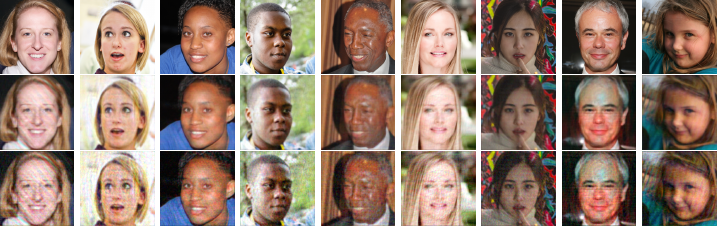}
    
\caption{Ground-truth test images (top row), reconstructions using the developed approach (middle row), and HIO initialization results (bottom row) for the case with parameters: $\alpha=3$, $N=1$, $\eta=0.78$, $\eta_b=0.17$, $t=30$, and $T_{init}=300$.}
\label{fig:fig4}
\end{figure}

For visual evaluation, sample reconstructions are also provided in Figures \ref{fig:fig1}, \ref{fig:fig2}, \ref{fig:fig3}, and \ref{fig:fig4}, together with the ground truth images. The results demonstrate that the integration of pretrained diffusion models with the HIO algorithm enables effective denoising and phase retrieval in many instances. However, although reduced, some HIO artifacts still exist in the final reconstructions. An important future work is to explore how to modify the update steps in Eq.~\eqref{ourApproach} to reduce these artifacts in the reconstructions 
as well as to extend the mathematical framework for the noisy case.

\subsection{Performance Analysis with Experimental Data}

We also evaluate the performance of the developed approach with experimental data to demonstrate its effectiveness in practical scenarios. For this, we consider the imaging application through random scattering media and apply the extended version of the approach in Section~\ref{section:extension} for the recovery of real-valued non-negative images. The chosen AP approach is again HIO since there are spatial-domain constraints to enforce. 

We use publicly available experimental data obtained with the setup described in \cite{metzler2017coherent}. Double phase retrieval methods were used to estimate the system (transmission) matrix $\mathbf{A}$. This calibration was performed using the prVAMP algorithm and amplitude-only spatial light modulator (SLM). 
The tested reconstruction methods are used with the estimated transmission matrix $\mathbf{A}$ from this calibration step. For DDRM-PR, the following hyperparameters are used: \( N = 1 \), \( \eta = 1.0 \), \( \eta_b = 0.0 \), \( t = 35 \), and \( T_{\text{init}} = 220 \).  

\begin{figure}[htb!]
    \centering

    \includegraphics[width=\columnwidth]{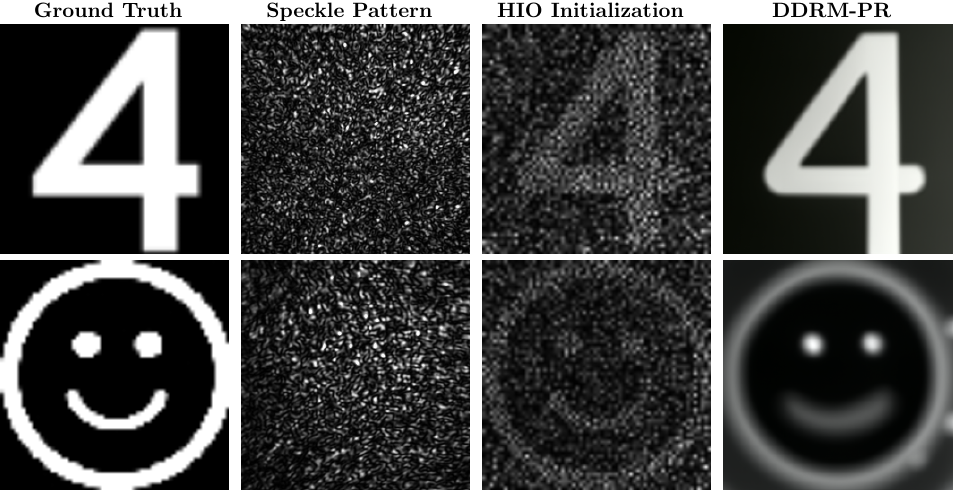}
    \caption{Performance comparison with experimental data. Each row corresponds to a different test image. The first column shows the ground truth representing the target amplitude-only signal. The second column displays the speckle pattern captured on the detector, which is the observed intensity measurement after passing through the scattering medium. The third column presents the HIO result illustrating the initial reconstruction obtained with HIO algorithm. The last column shows the reconstruction of DDRM-PR, which integrates pretrained diffusion models for enhanced phase retrieval.}
    \label{fig:comparisonEmpirical}
\end{figure}

Average reconstruction performance over the test dataset of \cite{metzler2017coherent} is provided in Table~\ref{table:tableempirical} for both DDRM-PR and HIO. For visual comparison, the reconstructions are also shown in Figure~\ref{fig:comparisonEmpirical}. Here the columns represent (from left to right) the ground truth, speckle patterns, HIO result (initialization), and DDRM-PR reconstructions. As seen, while HIO reconstructions are highly noisy, DDRM-PR 
significantly outperforms HIO both visually and quantitatively (in terms of all three metrics), 
with artifacts reduced and fine details preserved. In fact, the DDRM-PR reconstructions closely resemble the ground truth images, and structural and perceptual details are better preserved than HIO.

These results demonstrate the robustness and efficiency of our approach for real-world phase retrieval problems. Although the target images in this application have different characteristics than the images used in the training of the diffusion model, 
DDRM-PR is capable of providing a robust performance. In fact, the diffusion model used was pretrained on ImageNet, a dataset of natural RGB images. However, the experimental data used grayscale images which are black and white. Even under these large discrepancies, our approach produces high-quality reconstructions, showing its generalization capability and adaptability to different settings.

\begin{table}[htb!]
	\centering
	\caption{\bf The average reconstruction performance with experimental data}
    \small
	    \begin{adjustbox}{max width=\columnwidth}
     
		\begin{tabu}{@{\extracolsep{4pt}}cccc@{}}
			\hline
Methods &PSNR $\uparrow$ &SSIM $\uparrow$&LPIPS $\downarrow$
\\
\hline
HIO Stage&7.85&0.049&0.69\\
DDRM-PR&13.12&0.32&0.28\\
   
\hline					
	\end{tabu}
     \end{adjustbox}

\label{table:tableempirical}
\end{table}

While our approach has shown promising results, further improvements can be achieved by fine-tuning the pretrained diffusion models on datasets more closely matching the target image distribution. Additionally, as a future direction, one can leverage more recent diffusion models such as Stable Diffusion \cite{rombach2022high}, which exhibit better image generation performance. Such models trained on larger and more diverse datasets can 
be incorporated into our framework to further enhance the reconstruction quality and robustness of the approach, particularly when dealing with complex real-world data and broader range of image distributions.

Moreover, extending our experiments to address the recovery of complex-valued images presents an exciting avenue for future work. While the current experimental results focus on reconstructing an amplitude-only, real-valued signal, exploring the reconstruction of complex-valued signals (with both amplitude and phase components) would expand the applicability of our method to a wider array of real-world applications. These advancements would further validate the versatility and adaptability of the proposed framework in handling diverse phase retrieval challenges.

\section{Conclusion}
\label{sec:conclusion}

In this paper, we propose a new approach for the nonlinear phase retrieval problem by exploiting the efficient and unsupervised posterior sampling framework of Denoising Diffusion Restoration Models (DDRM). The approach integrates DDRM with the 
classical alternating projection (AP) methods to utilize pretrained diffusion priors for phase retrieval.
Exploiting pretrained unconditional diffusion models offer several practical advantages. Since the pretrained models are developed using large and diverse image datasets, they capture a wide range of features that are crucial for effective denoising and reconstruction. This integration also bypasses the need for retraining specific to the phase retrieval task, making the method more accessible and easier to implement in various settings. This flexibility sets our approach apart from many learning-based direct inversion methods, which require retraining for a different measurement setup.  

Our approach can be seen as a regularized extension of classical alternating projection methods, such as GS and HIO, with a diffusion-based prior. This synergy combines the strengths of traditional iterative algorithms and modern generative modeling techniques to enable more robust, fast, and high-quality reconstructions. Furthermore, the flexibility of the method allows it to adapt easily to different measurement types, such as Fourier magnitude, coded diffraction patterns, scattering, and Fourier ptychography. Although the results in this paper focus on reconstructing real-valued images, the proposed framework is equally applicable to the recovery of complex-valued signals. 

While our results obtained with experimental and simulated data demonstrate the potential of this approach for improving the AP methods, there are limitations and areas of improvement that present opportunities for future research. One key limitation is that the current formulation uses noiseless measurements. Future work could focus on extending the mathematical framework of DDRM to the noisy case for nonlinear problems and explicitly designing the sampling pipeline to handle observation noise. Exploring how to modify the update steps of the algorithm can help to further reduce the reconstruction artifacts. Another limitation arises due to the use of AP methods for initialization. When the AP reconstructions contain artifacts, these can persist in the final output after applying our pipeline. These artifacts, however, are typically localized in different regions of the image for different initialization runs. This suggests that combining multiple initializations by leveraging advanced pretrained diffusion models could improve reconstruction quality. For instance, diffusion models capable of conditioning on multiple image inputs could integrate information from various initializations to produce higher-quality reconstructions. 

As mentioned, a notable strength of our approach is its reliance on pretrained denoisers, which enhances generalization and adaptability across different phase retrieval problems. However, task-specific training could further improve performance. Pretrained models are effective for general applications, particularly when sufficient data for training is unavailable. However, in cases where data are abundant, training a diffusion model tailored to a specific phase retrieval problem could enable one to address problem-specific artifacts and intricacies. Future work could explore training diffusion models specifically tailored for phase retrieval tasks, as investigated in \cite{kaya2024data}.

Finally, incorporating more advanced pretrained diffusion models, such as those capable of conditional generation or trained on larger, more diverse datasets, could further enhance reconstruction performance. These models may provide improved robustness and adaptability, particularly for handling more difficult phase retrieval tasks. 

In conclusion, this work demonstrates the potential of integrating pretrained diffusion priors with the efficient AP methods for phase retrieval. The proposed framework provides a balance between generalization capability and 
reconstruction quality, offering a versatile solution that can be applied to a wide range of phase retrieval problems. By addressing the outlined limitations and exploring future directions, the approach can be further improved, paving the way for more advanced and robust phase retrieval methods.

\begin{backmatter}

\bmsection{Funding}
Türkiye Bilimsel ve Teknolojik Araştırma Kurumu (120E505).

\bmsection{Acknowledgment}
This study was funded in part by Scientific and Technological Research Council of Turkey (TUBITAK) under the Grant Number 120E505. Figen S. Oktem thanks TUBITAK for the support.

\bmsection{Disclosures}
The authors declare no conflicts of interest related to this
article.

\bmsection{Data availability}
Data underlying the results presented in this paper are not publicly available at this time but may be obtained from authors upon reasonable request. The source code is available at Ref.~\cite{ddrmpr_implementation}.

\end{backmatter}

\appendix
\section{Proof of Theorem 4.1}
\label{appendix:proofs}

We first provide some definitions and simple lemmas that will be used for the proof of Theorem 4.1.

\begin{definition}
\label{proof:defalphatsigmat}
Let $\alpha_t = \frac{1}{1 + \sigma_t^2}$ for all $t$. Equivalently,
\begin{equation}
\sigma_t = \sqrt{\frac{1 - \alpha_t}{\alpha_t}}, \quad \forall t.
\end{equation}
\end{definition}

\begin{definition}
\label{proof:defbluext}
Let \(\mathbf{\blue{x_t}} = \sqrt{\alpha_t} \mathbf{x_t}\), for all $t$. Cyan color will be used to denote this scaling.
\end{definition}

\begin{lemma}
\label{proof:lemmanoisecanfromdenoiser}
\begin{equation}
\frac{\blue{\mathbf{x_t}} - \sqrt{\alpha_t} \blue{\mathbf{x_0}}}{\sqrt{1 - \alpha_t}} \sim \mathcal{N}(\mathbf{0}, \mathbf{I}), \quad \forall t.
\end{equation}
\begin{proof}
From the assumed $q(\mathbf{x_t}, \mathbf{x_0})$, we know that
\begin{equation}
\frac{\mathbf{x_t} - \mathbf{x_0}}{\sigma_t} \sim \mathcal{N}(\mathbf{0}, \mathbf{I}).
\end{equation}
Using Definitions \ref{proof:defalphatsigmat} and \ref{proof:defbluext},
\begin{equation}
\frac{\mathbf{x_t} - \mathbf{x_0}}{\sigma_t} = \frac{\frac{\blue{\mathbf{x_t}}}{\sqrt{\alpha_t}}  - \frac{\blue{\mathbf{x_0}}}{\sqrt{\alpha_0}}}{\sqrt{\frac{1 - \alpha_t}{\alpha_t}}}.
\end{equation}
Since $\sigma_0$ is assumed to be $0$, $\alpha_0 = 1$, and, it follows that
\begin{equation}
\frac{\blue{\mathbf{x_t}} - \sqrt{\alpha_t} \blue{\mathbf{x_0}}}{\sqrt{1 - \alpha_t}} \sim \mathcal{N}(\mathbf{0}, \mathbf{I}).
\end{equation}
\end{proof}
\end{lemma}

\begin{corollary}
\label{proof:perfectestimatornormal}
    If we have a perfect 
    estimator of $\blue{\mathbf{x}_0}$ denoted by $\mathbf{x}_{\theta,t} = \mathbf{f}_{\theta}^{(t+1)}(\mathbf{x}_{t+1})$, then, 
\begin{equation}
\mathbf{\epsilon}_{\theta}^{(t+1)}(\mathbf{x}_{t+1}) = 
\frac{\blue{\mathbf{x_{t+1}}} - \sqrt{\alpha_{t+1}} \mathbf{x}_{\theta, t}}{\sqrt{1 - \alpha_{t+1}}} \sim \mathcal{N}(\mathbf{0}, \mathbf{I}).
\end{equation}
\end{corollary}

\begin{lemma}
\label{proof:lineartransformstats}
If $\mathbf{\eta} \sim \mathcal{N}(\mathbf{\mu}, \mathbf{\Sigma})$, then, $\mathbf{A} \mathbf{\eta} \sim \mathcal{N}(\mathbf{A} \mathbf{\mu}, \mathbf{A \Sigma A^T})$.
\end{lemma}

\begin{corollary}
\label{proof:linearorthotransformstatsnoise}
If $\mathbf{\eta} \sim \mathcal{N}(\mathbf{0}, \mathbf{I})$, then, $\mathbf{V} \mathbf{\eta} \sim \mathcal{N}(\mathbf{0}, \mathbf{I})$ for an orthogonal matrix $\mathbf{V}$.
\end{corollary}

\begin{lemma}[Reparametrization trick]
\label{proof:reparamtrick}
    If $w \sim \mathcal{N}(\mu, \sigma^2)$, then, we can write it as
    \begin{equation}
    w = \mu + \sigma \epsilon \quad \text{where} \quad \epsilon \sim \mathcal{N}(0, 1).
    \end{equation}
\end{lemma}

\begin{lemma}
\label{proof:HHSS}
    \begin{equation}
    \boldsymbol{H^\dagger H}
    = (\mathbf{V \Sigma^\dagger U^T})(\mathbf{U \Sigma V^T})
    = \mathbf{V \Sigma^\dagger}\mathbf{\Sigma V^T}
    = \mathbf{\Sigma^\dagger \Sigma}\mathbf{V V^T}
    = \mathbf{\Sigma^\dagger \Sigma}
    \end{equation}
    \begin{proof}
        Matrix multiplication with a square diagonal matrix is commutative.
    \end{proof}
\end{lemma}

To prove Theorem 4.1, we start with the original form of DDRM in the noiseless case:

\begin{equation}
\begin{aligned}
p_\theta^{(t)}\left(\overline{\mathbf{x}}_t^{(i)} \mid \mathbf{x}_{t+1}, \mathbf{y}\right)= & \begin{cases}\mathcal{N}\left(\overline{\mathbf{x}}_{\theta, t}^{(i)}+\sqrt{1-\eta^2} \sigma_t \frac{\overline{\mathbf{x}}_{t+1}^{(i)}-\overline{\mathbf{x}}_{\theta, t}^{(i)}}{\sigma_{t+1}}, \eta^2 \sigma_t^2\right) \\ \hfill \text { if } s_i=0 \\[10pt]
\mathcal{N}\left(\left(1-\eta_b\right) \overline{\mathbf{x}}_{\theta, t}^{(i)}+\eta_b \overline{\mathbf{y}}^{(i)}, \sigma_t^2 \right) \\ \hfill \text { otherwise }\end{cases}
\end{aligned}
\end{equation}

Using the reparametrization trick given in Lemma \ref{proof:reparamtrick}, we obtain
\begin{equation}
\begin{aligned}
\overline{\mathbf{x}}_t^{(i)} = & \begin{cases} \overline{\mathbf{x}}_{\theta, t}^{(i)}+\sqrt{1-\eta^2} \sigma_t \frac{\overline{\mathbf{x}}_{t+1}^{(i)}-\overline{\mathbf{x}}_{\theta, t}^{(i)}}{\sigma_{t+1}} + \sqrt{ \eta^2 \sigma_t^2 } \mathbf{\epsilon}_t^{(i)} & \text { if } s_i=0 \\
\left(1-\eta_b\right) \overline{\mathbf{x}}_{\theta, t}^{(i)}+\eta_b \overline{\mathbf{y}}^{(i)} + \sqrt{\sigma_t^2} \mathbf{\epsilon}_t^{'(i)}  & \text { otherwise }\end{cases}
\end{aligned}
\end{equation}
where $\mathbf{\epsilon}_t', \mathbf{\epsilon}_t \sim \mathcal{N}(\mathbf{0}, \mathbf{I})$.

Note that the matrix $\boldsymbol{\Sigma}^{\dagger} \boldsymbol{\Sigma}$ is a diagonal matrix with zeros at positions corresponding to zero singular values and ones elsewhere. This allows us to express $\overline{\mathbf{x}}_t$ in a more compact form:
\begin{equation}
\begin{aligned}
\overline{\mathbf{x}}_t = 
(\mathbf{I} - \boldsymbol{\Sigma}^{\dagger} \boldsymbol{\Sigma})&
\left(
\overline{\mathbf{x}}_{\theta, t}+\sqrt{1-\eta^2} \sigma_t \frac{\overline{\mathbf{x}}_{t+1}-\overline{\mathbf{x}}_{\theta, t}}{\sigma_{t+1}} + \eta \sigma_t \mathbf{\epsilon}_t 
\right)\\
+
\boldsymbol{\Sigma}^{\dagger} \boldsymbol{\Sigma}&
\left(
\left(1-\eta_b\right) \overline{\mathbf{x}}_{\theta, t}+\eta_b \overline{\mathbf{y}} + \sigma_t \mathbf{\epsilon}_t^{'}
\right)
\end{aligned}
\end{equation}

Replacing this with the spectral definition given in Definition \eqref{transformation1} yields
\begin{equation}
\begin{aligned}
\mathbf{V^T}{\mathbf{x}}_t\\
=
(\mathbf{I} - \boldsymbol{\Sigma}^{\dagger} \boldsymbol{\Sigma})&
\left(
\mathbf{V^T}{\mathbf{x}}_{\theta, t}+\sqrt{1-\eta^2} \sigma_t \frac{\mathbf{V^T}{\mathbf{x}}_{t+1}-\mathbf{V^T}{\mathbf{x}}_{\theta, t}}{\sigma_{t+1}} + \eta \sigma_t \mathbf{\epsilon}_t 
\right)\\
+  
\boldsymbol{\Sigma}^{\dagger} \boldsymbol{\Sigma}&
\left(
\left(1-\eta_b\right) \mathbf{V^T}{\mathbf{x}}_{\theta, t}+\eta_b \mathbf{\Sigma^\dagger U^T}{\mathbf{y}} + \sigma_t \mathbf{\epsilon}_t^{'}
\right)
\end{aligned}
\end{equation}

By multiplying both sides with $\mathbf{V}$, and recalling that multiplication with square diagonal matrices is commutative, we obtain
\begin{equation}
\begin{aligned}
{\mathbf{x}}_t = 
(\mathbf{I} - \boldsymbol{\Sigma}^{\dagger} \boldsymbol{\Sigma})&
\left(
{\mathbf{x}}_{\theta, t}+\sqrt{1-\eta^2} \sigma_t \frac{{\mathbf{x}}_{t+1}-{\mathbf{x}}_{\theta, t}}{\sigma_{t+1}} + \eta \sigma_t \mathbf{V}\boldsymbol{\epsilon}_t 
\right)\\
+
\boldsymbol{\Sigma}^{\dagger} \boldsymbol{\Sigma}&
\left(
\left(1-\eta_b\right) {\mathbf{x}}_{\theta, t}+\eta_b \boldsymbol{H^\dagger}{\mathbf{y}} + \sigma_t \mathbf{V} \boldsymbol{\epsilon}_t^{'}
\right)
\end{aligned}
\end{equation}

Using Lemma \ref{proof:HHSS} ($\boldsymbol{H^\dagger H} = \mathbf{\Sigma^\dagger \Sigma}$) and Corollary \ref{proof:linearorthotransformstatsnoise} ($\mathbf{V} \boldsymbol{\epsilon}_t = \mathbf{\epsilon}_t$), this becomes
\begin{equation}
\begin{aligned}
{\mathbf{x}}_t = 
(\mathbf{I} - \boldsymbol{H}^{\dagger} \boldsymbol{H})&
\left(
{\mathbf{x}}_{\theta, t}+\sqrt{1-\eta^2} \sigma_t \frac{{\mathbf{x}}_{t+1}-{\mathbf{x}}_{\theta, t}}{\sigma_{t+1}} + \eta \sigma_t \mathbf{\epsilon}_t 
\right)\\
+
\boldsymbol{H}^{\dagger} \boldsymbol{H}&
\left(
\left(1-\eta_b\right) {\mathbf{x}}_{\theta, t}+\eta_b \boldsymbol{H^\dagger}{\mathbf{y}} + \sigma_t \mathbf{\epsilon}_t^{'}
\right)
\end{aligned}
\end{equation}

Using Definitions \ref{proof:defbluext} and \ref{proof:defalphatsigmat}:
\begin{equation}
\begin{aligned}
\frac{\blue{\mathbf{x}_t}}{\sqrt{\alpha_t}} =\\ 
(\mathbf{I} - \boldsymbol{H}^{\dagger} \boldsymbol{H})&
\left(
{\mathbf{x}}_{\theta, t}+\sqrt{1-\eta^2} \sqrt{\frac{1 - \alpha_t}{\alpha_t}} \frac{\frac{\blue{\mathbf{x}_{t+1}}}{\sqrt{\alpha_{t+1}}}-{\mathbf{x}}_{\theta, t}}{\sqrt{\frac{1 - \alpha_{t+1}}{\alpha_{t+1}}}} + \eta \sqrt{\frac{1 - \alpha_t}{\alpha_t}} \mathbf{\epsilon}_t 
\right)\\
+
\boldsymbol{H}^{\dagger} \boldsymbol{H}&
\left(
\left(1-\eta_b\right) {\mathbf{x}}_{\theta, t}+\eta_b \boldsymbol{H^\dagger}{\mathbf{y}} + \sqrt{\frac{1 - \alpha_t}{\alpha_t}} \mathbf{\epsilon}_t^{'}
\right)
\end{aligned}
\end{equation}

This can be simplified as
\begin{equation}
\begin{aligned}
\blue{\mathbf{x}_t} = 
(\mathbf{I} - \boldsymbol{H}^{\dagger} \boldsymbol{H})&
(\sqrt{\alpha_t}
{\mathbf{x}}_{\theta, t}+\sqrt{1-\eta^2} \sqrt{1 - \alpha_t} \frac{\blue{\mathbf{x}_{t+1}}-\sqrt{\alpha_{t+1}}{\mathbf{x}}_{\theta, t}}{\sqrt{{1 - \alpha_{t+1}}}} \\ &+ \eta \sqrt{1 - \alpha_t} \mathbf{\epsilon}_t 
)\\
+
\boldsymbol{H}^{\dagger} \boldsymbol{H}&
\left(
\sqrt{\alpha_t} \left(1-\eta_b\right) {\mathbf{x}}_{\theta, t}+\sqrt{\alpha_t} \eta_b \boldsymbol{H^\dagger}{\mathbf{y}} + \sqrt{1 - \alpha_t} \mathbf{\epsilon}_t^{'}
\right)
\end{aligned}
\end{equation}

Rearranging the terms and utilizing the property $\boldsymbol{H} \boldsymbol{H}^\dagger \boldsymbol{H} = \boldsymbol{H}$ leads to the following expression:
\begin{equation}
\begin{aligned}
\blue{\mathbf{x}_t} =&
\sqrt{\alpha_t}
{\mathbf{x}}_{\theta, t} + \sqrt{\alpha_t} (-1 + (1-\eta_b)) \boldsymbol{H}^{\dagger} \boldsymbol{H} {\mathbf{x}}_{\theta, t} + \sqrt{\alpha_t} \eta_b \boldsymbol{H^\dagger}{\mathbf{y}}\\
&+ \eta \sqrt{1 - \alpha_t} \mathbf{\epsilon}_t \\
&+ \boldsymbol{H}^{\dagger} \boldsymbol{H} (-\eta + 1) \sqrt{1 - \alpha_t} \mathbf{\epsilon}_t' \\
&+ (\mathbf{I} - \boldsymbol{H}^{\dagger} \boldsymbol{H}) \sqrt{1-\eta^2} \sqrt{1 - \alpha_t} \frac{\blue{\mathbf{x}_{t+1}}-\sqrt{\alpha_{t+1}}{\mathbf{x}}_{\theta, t}}{\sqrt{{1 - \alpha_{t+1}}}} \\
\end{aligned}
\end{equation}

Using Corollary \ref{proof:perfectestimatornormal} and approximating $\sqrt{1-\eta^2} \approx 1 - \eta$ for $\eta \in [0, 1]$, we obtain the following final form:
\begin{equation}
\begin{aligned}
\blue{\mathbf{x}_t} =&
\sqrt{\alpha_t} \left(
{\mathbf{x}}_{\theta, t} +  \eta_b \left({\mathbf{x}}_{\theta, t} -\boldsymbol{H}^{\dagger} \boldsymbol{H} {\mathbf{x}}_{\theta, t} +  \boldsymbol{H^\dagger}{\mathbf{y}} \right) - \eta_b {\mathbf{x}}_{\theta, t} \right)\\
&+  \sqrt{1 - \alpha_t} \eta \mathbf{\epsilon}_t \\
&+ \boldsymbol{H}^{\dagger} \boldsymbol{H} (1 - \eta) \sqrt{1 - \alpha_t} \epsilon_\theta^{(t+1)}\left(\mathbf{x}_{t+1}\right) \\
&+ (\mathbf{I} - \boldsymbol{H}^{\dagger} \boldsymbol{H}) {(1-\eta)} \sqrt{1 - \alpha_t} \epsilon_\theta^{(t+1)}\left(\mathbf{x}_{t+1}\right) \\
=&
\sqrt{\alpha_t} \left(
 \eta_b \left({\mathbf{x}}_{\theta, t} -\boldsymbol{H}^{\dagger} \boldsymbol{H} {\mathbf{x}}_{\theta, t} +  \boldsymbol{H^\dagger}{\mathbf{y}} \right) + {(1-\eta_b) \mathbf{x}}_{\theta, t} \right)\\
&+  \sqrt{1 - \alpha_t} \left( \eta \mathbf{\epsilon}_t +
 (1 - \eta)  \mathbf{\epsilon}_\theta^{(t+1)}\left(\mathbf{x}_{t+1}\right) \right)
\end{aligned}
\end{equation}

\bibliography{references}

\begin{thebibliography}{10}
\newcommand{\enquote}[1]{``#1''}

\bibitem{Dong_2023}
J.~Dong, L.~Valzania, A.~Maillard, T.-a. Pham, S.~Gigan, and M.~Unser, \enquote{Phase retrieval: From computational imaging to machine learning: A tutorial,} {\protect\JournalTitle{IEEE Signal Processing Magazine}} \textbf{40}, 45–57 (2023).

\bibitem{shechtman2015phase}
Y.~Shechtman, Y.~C. Eldar, O.~Cohen, H.~N. Chapman, J.~Miao, and M.~Segev, \enquote{Phase retrieval with application to optical imaging: a contemporary overview,} {\protect\JournalTitle{IEEE Signal Processing Magazine}} \textbf{32}, 87--109 (2015).

\bibitem{Fannjiang2020TheNO}
A.~Fannjiang and T.~Strohmer, \enquote{The numerics of phase retrieval,} {\protect\JournalTitle{Acta Numerica}} \textbf{29}, 125 -- 228 (2020).

\bibitem{stefanoqianpty}
J.~Qian, C.~Yang, A.~Schirotzek, F.~Maia, and S.~Marchesini, \enquote{Efficient algorithms for ptychographic phase retrieval, in inverse problems and applications,} {\protect\JournalTitle{Contemp. Math}} \textbf{615}, 261--280 (2014).

\bibitem{Maiden:17}
A.~Maiden, D.~Johnson, and P.~Li, \enquote{Further improvements to the ptychographical iterative engine,} {\protect\JournalTitle{Optica}} \textbf{4}, 736--745 (2017).

\bibitem{Sinha:17}
A.~Sinha, J.~Lee, S.~Li, and G.~Barbastathis, \enquote{Lensless computational imaging through deep learning,} {\protect\JournalTitle{Optica}} \textbf{4}, 1117--1125 (2017).

\bibitem{jin2017deep}
K.~H. Jin, M.~T. McCann, E.~Froustey, and M.~Unser, \enquote{Deep convolutional neural network for inverse problems in imaging,} {\protect\JournalTitle{IEEE Transactions on Image Processing}} \textbf{26}, 4509--4522 (2017).

\bibitem{Isil:19}
\c{C}. I\c{s}{ı}l, F.~S. Oktem, and A.~Ko\c{c}, \enquote{Deep iterative reconstruction for phase retrieval,} {\protect\JournalTitle{Appl. Opt.}} \textbf{58}, 5422--5431 (2019).

\bibitem{venkatakrishnan2013plug}
S.~V. Venkatakrishnan, C.~A. Bouman, and B.~Wohlberg, \enquote{Plug-and-play priors for model based reconstruction,} in \emph{IEEE Global Conference on Signal and Information Processing,}  (IEEE, 2013), pp. 945--948.

\bibitem{chan2017}
S.~H. Chan, X.~Wang, and O.~A. Elgendy, \enquote{Plug-and-play {ADMM} for image restoration: Fixed-point convergence and applications,} {\protect\JournalTitle{IEEE Transactions on Computational Imaging}} \textbf{3}, 84--98 (2017).

\bibitem{romano2017}
Y.~Romano, M.~Elad, and P.~Milanfar, \enquote{The little engine that could: Regularization by denoising {(RED)},} {\protect\JournalTitle{SIAM Journal on Imaging Sciences}} \textbf{10}, 1804--1844 (2017).

\bibitem{pmlr-v80-metzler18a}
C.~Metzler, P.~Schniter, A.~Veeraraghavan, and R.~Baraniuk, \enquote{pr{D}eep: Robust phase retrieval with a flexible deep network,} in \emph{International Conference on Machine Learning,}  (2018), pp. 3498--3507.

\bibitem{Isil:20}
\c{C}. I\c{s}{ı}l and F.~S. Oktem, \enquote{Model-based phase retrieval with deep denoiser prior,} in \emph{Imaging and Applied Optics Congress,}  (Optica Publishing Group, 2020), p. CF2C.5.

\bibitem{isil2024}
\c{C}. I\c{s}{ı}l and F.~S. Oktem, \enquote{A deep plug-and-play approach for phase retrieval,} {\protect\JournalTitle{arXiv preprint arXiv:2411.18967}}  (2024).

\bibitem{Wei2020TuningfreePP}
K.~Wei, A.~I. Avil{\'e}s-Rivero, J.~Liang, Y.~Fu, C.-B. Sch{\"o}nlieb, and H.~Huang, \enquote{Tuning-free plug-and-play proximal algorithm for inverse imaging problems,} in \emph{International Conference on Machine Learning,}  (2020).

\bibitem{Wang_2024}
K.~Wang, L.~Song, C.~Wang, Z.~Ren, G.~Zhao, J.~Dou, J.~Di, G.~Barbastathis, R.~Zhou, J.~Zhao, and E.~Y. Lam, \enquote{On the use of deep learning for phase recovery,} {\protect\JournalTitle{Light: Science \& Applications}} \textbf{13} (2024).

\bibitem{kawar2022denoising}
B.~Kawar, M.~Elad, S.~Ermon, and J.~Song, \enquote{Denoising diffusion restoration models,} in \emph{Advances in Neural Information Processing Systems,}  (2022).

\bibitem{dhariwal2021diffusion}
P.~Dhariwal and A.~Nichol, \enquote{Diffusion models beat {GANs} on image synthesis,} in \emph{Advances in Neural Information Processing Systems,}  (2021), pp. 8780--8794.

\bibitem{Kawar2022JPEGAC}
B.~Kawar, J.~Song, S.~Ermon, and M.~Elad, \enquote{{JPEG} artifact correction using denoising diffusion restoration models,} in \emph{Neural Information Processing Systems (NeurIPS) Workshop on Score-Based Methods,}  (2022).

\bibitem{hayes1982}
M.~Hayes, \enquote{The reconstruction of a multidimensional sequence from the phase or magnitude of its {Fourier} transform,} {\protect\JournalTitle{IEEE Transactions on Acoustics, Speech, and Signal Processing}} \textbf{30}, 140--154 (1982).

\bibitem{Schulz1992image}
T.~J. Schulz and D.~L. Snyder, \enquote{Image recovery from correlations,} {\protect\JournalTitle{J. Opt. Soc. Am. A}} \textbf{9}, 1266--1272 (1992).

\bibitem{1987_fienup_pr_astronomy}
J.~C.~Dainty and J.~Fienup, \enquote{Phase retrieval and image reconstruction for astronomy,} {\protect\JournalTitle{Image Recovery: Theory Appl}} \textbf{13}, 231--275 (1987).

\bibitem{10.1117/12.7976761}
J.~R. Fienup and P.~S. Idell, \enquote{{Imaging Correlography With Sparse Arrays Of Detectors},} {\protect\JournalTitle{Optical Engineering}} \textbf{27}, 279778 (1988).

\bibitem{katz2014non}
O.~Katz, P.~Heidmann, M.~Fink, and S.~Gigan, \enquote{Non-invasive single-shot imaging through scattering layers and around corners via speckle correlations,} {\protect\JournalTitle{Nature photonics}} \textbf{8}, 784--790 (2014).

\bibitem{miao2008extending}
J.~Miao, T.~Ishikawa, Q.~Shen, and T.~Earnest, \enquote{Extending x-ray crystallography to allow the imaging of noncrystalline materials, cells, and single protein complexes,} {\protect\JournalTitle{Annu. Rev. Phys. Chem.}} \textbf{59}, 387--410 (2008).

\bibitem{gs1978}
R.~W. Gerchberg and W.~O. Saxton, \enquote{A practical algorithm for the determination of phase from image and diffraction plane pictures,} {\protect\JournalTitle{Optik}} \textbf{35}, 237--250 (1972).

\bibitem{fienup1978reconstruction}
J.~R. Fienup, \enquote{Reconstruction of an object from the modulus of its {Fourier} transform,} {\protect\JournalTitle{Optics Letters}} \textbf{3}, 27--29 (1978).

\bibitem{fienup1982comparison}
J.~R. Fienup, \enquote{Phase retrieval algorithms: a comparison,} {\protect\JournalTitle{Appl. Opt.}} \textbf{21}, 2758--2769 (1982).

\bibitem{marchesini2007invited}
S.~Marchesini, \enquote{Invited article: A unified evaluation of iterative projection algorithms for phase retrieval,} {\protect\JournalTitle{Review of Scientific Instruments}} \textbf{78} (2007).

\bibitem{li2023diffusion}
X.~Li, Y.~Ren, X.~Jin, C.~Lan, X.~Wang, W.~Zeng, X.~Wang, and Z.~Chen, \enquote{Diffusion models for image restoration and enhancement--a comprehensive survey,} {\protect\JournalTitle{arXiv preprint arXiv:2308.09388}}  (2023).

\bibitem{ho2020denoising}
J.~Ho, A.~Jain, and P.~Abbeel, \enquote{Denoising diffusion probabilistic models,} {\protect\JournalTitle{Advances in neural information processing systems}} \textbf{33}, 6840--6851 (2020).

\bibitem{saharia2022photorealistic}
C.~Saharia, W.~Chan, S.~Saxena, L.~Li, J.~Whang, E.~L. Denton, K.~Ghasemipour, R.~Gontijo~Lopes, B.~Karagol~Ayan, T.~Salimans \emph{et~al.}, \enquote{Photorealistic text-to-image diffusion models with deep language understanding,} {\protect\JournalTitle{Advances in neural information processing systems}} \textbf{35}, 36479--36494 (2022).

\bibitem{rombach2022high}
R.~Rombach, A.~Blattmann, D.~Lorenz, P.~Esser, and B.~Ommer, \enquote{High-resolution image synthesis with latent diffusion models,} in \emph{Proceedings of the IEEE/CVF conference on computer vision and pattern recognition,}  (2022), pp. 10684--10695.

\bibitem{netrapalli2013phase}
P.~Netrapalli, P.~Jain, and S.~Sanghavi, \enquote{Phase retrieval using alternating minimization,} {\protect\JournalTitle{Advances in Neural Information Processing Systems}} \textbf{26} (2013).

\bibitem{metzler2017coherent}
C.~A. Metzler, M.~K. Sharma, S.~Nagesh, R.~G. Baraniuk, O.~Cossairt, and A.~Veeraraghavan, \enquote{Coherent inverse scattering via transmission matrices: Efficient phase retrieval algorithms and a public dataset,} in \emph{2017 IEEE International Conference on Computational Photography (ICCP),}  (IEEE, 2017), pp. 1--16.

\bibitem{chandra2019phasepack}
R.~Chandra, T.~Goldstein, and C.~Studer, \enquote{Phasepack: A phase retrieval library,} in \emph{2019 13th International conference on Sampling Theory and Applications (SampTA),}  (IEEE, 2019), pp. 1--5.

\bibitem{shewchuk1994introduction}
J.~Shewchuk, \emph{An Introduction to the Conjugate Gradient Method Without the Agonizing Pain} (Carnegie-Mellon University. Department of Computer Science, 1994).

\bibitem{lee2022complex}
C.~Lee, H.~Hasegawa, and S.~Gao, \enquote{Complex-valued neural networks: A comprehensive survey,} {\protect\JournalTitle{IEEE/CAA Journal of Automatica Sinica}} \textbf{9}, 1406--1426 (2022).

\bibitem{bassey2021survey}
J.~Bassey, L.~Qian, and X.~Li, \enquote{A survey of complex-valued neural networks,} {\protect\JournalTitle{arXiv preprint arXiv:2101.12249}}  (2021).

\bibitem{choi2018phase}
H.-S. Choi, J.-H. Kim, J.~Huh, A.~Kim, J.-W. Ha, and K.~Lee, \enquote{Phase-aware speech enhancement with deep complex u-net,} in \emph{International Conference on Learning Representations,}  (2018).

\bibitem{karras2018}
T.~Karras, T.~Aila, S.~Laine, and J.~Lehtinen, \enquote{Progressive growing of {GANs} for improved quality, stability, and variation,} in \emph{International Conference on Learning Representations (ICLR),}  (2018).

\bibitem{wang2004image}
Z.~Wang, A.~C. Bovik, H.~R. Sheikh, and E.~P. Simoncelli, \enquote{Image quality assessment: from error visibility to structural similarity,} {\protect\JournalTitle{IEEE Transactions on Image Processing}} \textbf{13}, 600--612 (2004).

\bibitem{zhang2018unreasonable}
R.~Zhang, P.~Isola, A.~A. Efros, E.~Shechtman, and O.~Wang, \enquote{The unreasonable effectiveness of deep features as a perceptual metric,} in \emph{IEEE Conference on Computer Vision and Pattern Recognition,}  (2018), pp. 586--595.

\bibitem{kaya2024data}
M.~O. Kaya, \enquote{Data-driven phase retrieval using deep generative models,} Master's thesis, Middle East Technical University (Turkey) (2024).

\bibitem{ddrmpr_implementation}
M.~O. Kaya, \enquote{{DDRM-PR}: Official implementation of {DDRM} phase retrieval paper,} \url{https://github.com/METU-SPACE-Lab/ddrm-pr} (2024). GitHub repository.

\end{thebibliography}

\end{document}